\documentclass[twocolumn,trackchanges]{aastex631}
\usepackage{amsmath}
\usepackage{comment}

\begin{document}

\title{Exploring filament galaxies using {\em AstroSat}/ UVIT}

\correspondingauthor{Divya Pandey}
\email{divyapandey1212@gmail.com}

\author{Divya Pandey}
\affiliation{Aryabhatta Research Institute of Observational Sciences, Manora Peak, Nainital 263 002, India}
\affiliation{Department of Physics and Astronomy, National Institute of Technology, Rourkela, Odisha 769 008, India}
\author{Kanak Saha}
\affiliation{Inter-University Centre for Astronomy \& Astrophysics, Postbag 4, Ganeshkhind, Pune 411 007, India}


\author{Ananta C. Pradhan}
\affiliation{Department of Physics and Astronomy, National Institute of Technology, Rourkela, Odisha 769 008, India}



\begin{abstract}
We present results from our deep Far-ultraviolet (FUV) survey using {\em AstroSat}/UVIT of a filamentary structure at $z$ $\sim$ $0.072$. A total of four filaments comprising 58 galaxies were probed in our study. We detect 18 filament galaxies in our FUV observation. All filament galaxies are further classified based on their photometric color, nuclear activity, and morphology. The filaments contain galaxies with mixed stellar population types and structures. We do not detect galaxies in our UVIT survey up to a distance of 0.4~Mpc $h^{-1}$ from the filament axis, implying a {lack} of recent star formation in the inner region of filaments. {The FUV star formation rate (SFR) for star-forming galaxies agrees well with the SFR$_{\rm 144 MHz}$ calculated using Low-Frequency Array (LOFAR) radio-continuum observations.} We witness an increase in FUV specific-SFR (sSFR) of filament galaxies with increasing distance from the filament spine (D$_{\rm fil}$). The intermediate-to-high stellar mass filament galaxies were more star-forming than cluster galaxies in a fixed stellar mass bin. The FUV morphology of some filament galaxies detected in the filament outskirts (D$_{\rm fil}$ $\gtrsim$ 0.7~Mpc $h^{-1}$) is comparable to or slightly extended than their optical counterpart. The mass assembly of galaxies examined by estimating $(FUV-r)$ color gradients show that more `red-cored' galaxies reside in the outer region of the filaments. Our results prove that the likelihood of merger interaction and gas starvation increases when approaching the filament spine. We report a definitive and in-homogeneous impact of filaments on the galaxies residing inside them.
\end{abstract}



\section{Introduction} \label{sec:intro}
{Early redshift surveys \citep[e.g.][]{1986Delapp, 2001Colless} and numerical simulations \citep[e.g.][]{1985Efstathiou} reveal the large-scale structure (LSS), comprising clusters, voids, walls and filaments \citep{1996Bond}. The filaments are the most dominant LSS as} around 50\% of the total mass fraction in the cosmic web resides within filaments, while the LSS occupies only $\sim$5\% of the cosmic volume \citep{2014Cautun}. Filaments inhabit the spaces between massive clusters and inside voids \citep{2014Tempel}. The LSS is not just an assembly of galaxies; it provides a dynamic pathway in channeling the galaxies to massive clusters. Apart from the cool baryonic matter in galaxies, filaments are shown to be the carrier of warm-hot intergalactic matter \citep[WHIM;][]{2021TNG_WHIM}. The filamentary pattern in the cosmic web is well-illustrated both in simulations \citep{2005Colberg, 2014Cautun} as well as observations \citep[e.g.,][]{2003Durret, 2005Potter}. The detection of filaments within the cosmic web is an intricate task due to their {variable thickness depending on the surrounding} environment density. There are various algorithms available for extracting filaments in the cosmic web, such as Bisous \citep{2014Tempel}, DisPerSE \citep{2011sousbie}, NEXUS+ \citep{2013Cautun,2014Cautun}.


Several large-scale phenomena, such as cosmic web enhancement and detachment, are reported to occur inside filaments. A few filament galaxies are shown to host ionized gas clouds in their circumgalactic region due to cosmic web enhancement \citep{2019Vulcani}. On the other hand, it is theoretically indicated that galaxies experience gas starvation caused by cosmic web detachment inside filaments due to the non-linear interaction of the cosmic web \citep{2019calvo}.{ The large-scale matter distribution along filament is found to be highly non-uniform \citep{2000Lee, Tempel_2013, 2014Cautun}, depending on where the filaments originated. Detailed study of the cosmic structure shows that the thickness of filaments, as well as the angle between galaxy halo spin and filament spine, profoundly impact the properties of resident galaxies \citep{2018GaneshaiahVeena, 2019GaneshaiahVeena, 2021GaneshaiahVeena}.}



There are several observation- \citep[e.g.,][]{Tempel_2013,2017Kuutma, 2018mahajan,2021Lee} and simulation-based \citep[e.g.,][]{2018GaneshaiahVeena, 2019GaneshaiahVeena,2020Singh, 2023Hasan} studies performed to probe the structural and physical properties of filament galaxies. Most of these analyses are based on optical imaging and spectroscopic datasets. {A few factors, such as the local environment density, the distance of galaxies from the filament spine, and the angle between galaxy halo spin alignment and filament axis, impact the galaxy evolution inside filaments \citep{2020Bonjean, 2018GaneshaiahVeena, 2021GaneshaiahVeena}.} Studies focusing on {characterizing} the properties of galaxies as a function of their distances from the spine of filaments found that the passive galaxy fraction increases closer to the filament axis \citep[]{2017Kuutma, 2018Laigle, 2020Singh, 2020Bonjean}. The authors report a morphological transformation of filament galaxies from late- to early-type, followed by a suppression in ongoing star formation when approaching the filament spine from the void environment.

{The spin alignment of galaxies in filaments is another determinant of the properties of galaxies. \citet{Tempel_2013} presents strong observational evidence suggesting that the spins of spiral galaxies align parallel to the filament axis, whereas the minor axis of elliptical galaxies typically orient themselves perpendicular to the filament axis. \citet{2018GaneshaiahVeena, 2019GaneshaiahVeena} highlight mass dependence on the halo spin and filament alignment. The authors show that low-mass halos tend to reside parallel to the filament axis and high-mass halos align preferentially perpendicular to the filament. These observations are a manifestation of large-scale tidal fields explained by the tidal torque theory \citep[TTT;][]{hoyle1949proceedings, 1969Peebles, 1984White}. The tidal fields affect gas inflow history inside filaments, impacting the ongoing star formation.}



The ultraviolet (UV) emission broadly divided into two wavelength ranges, far-UV (FUV) and near-UV (NUV), from galaxies is an ideal tracer of recent star formation in galaxies \citep{2007Gdp}. The average timescale of FUV emission is roughly 0.1 Gyr, even shorter than the NUV emission timescale of about one Gyr. A few investigations in the past have probed the properties of filament galaxies with the help of UV and UV-optical colors \citep[e.g.,][]{2016Alpaslan, 2018mahajan, 2021Lee}. The authors examine the gradient in UV colors, NUV star formation rates (SFRs), and stellar mass of filament galaxies when approaching the filament spine. These analyses hinted towards a definitive transformation in UV properties along the filament. Most of these investigations are based on the observations from {\em Galaxy Evolution Explorer \citep[ GALEX,][]{2007Morrisey}} which has performed a shallow survey of the entire sky having a typical exposure time of 100 s with a resolution of $\approx$5$^{\prime\prime}$ \citep{2007Morrisey}. It is worth noting that a deep and resolved FUV survey of the filament environment would be helpful to trace star formation activity in filaments even at the faint magnitude limit.
 
\begin{figure*}
    \centering
    \includegraphics[width=0.7\linewidth]{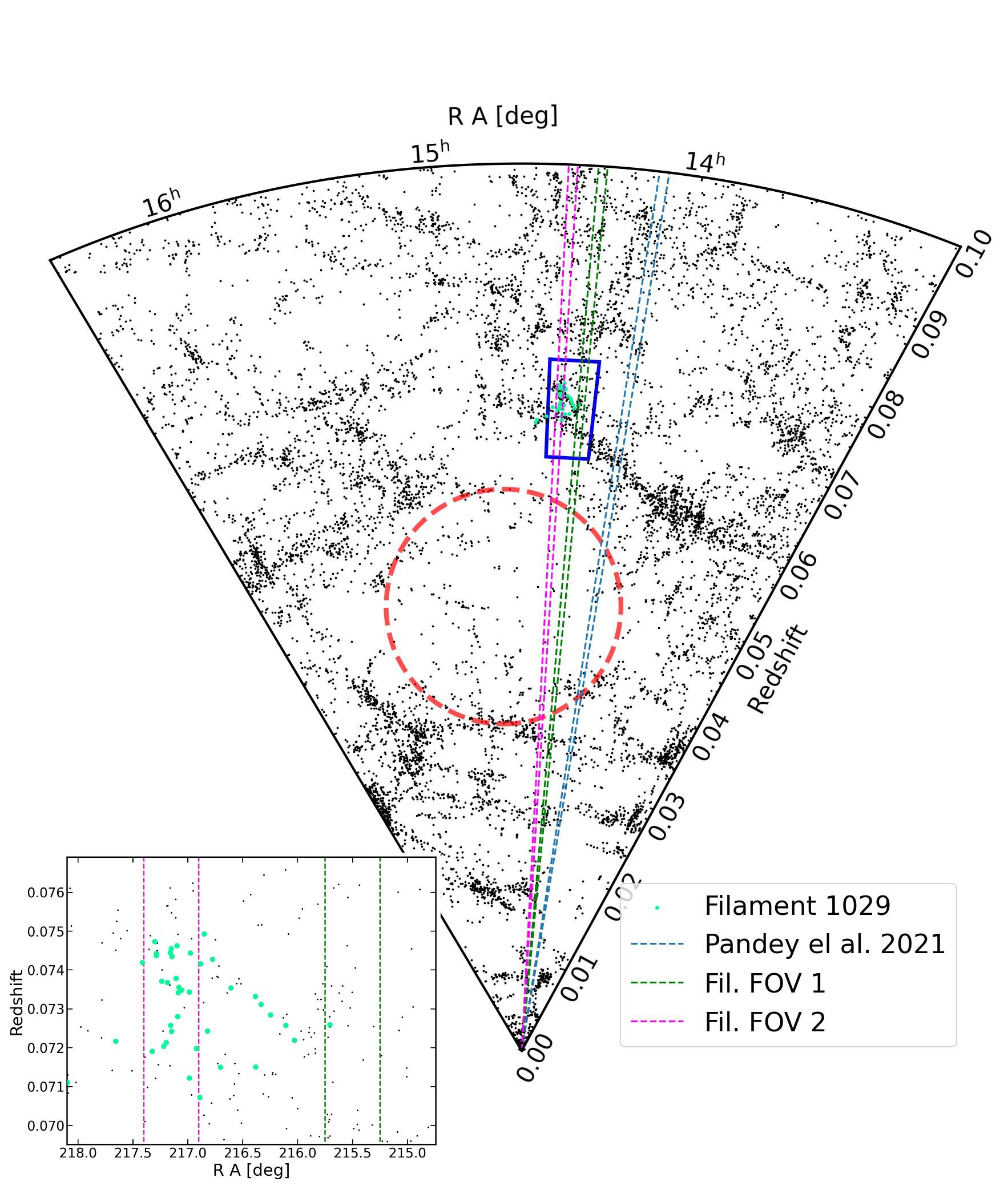}
    
    \caption{Redshift cone diagram showing the distribution of galaxies in the line-of-sight of UVIT observation. Each black dot in the figure represents a galaxy taken from {\em SDSS} DR 16. A blue rectangular box surrounds the filamentary structure observed in our UVIT survey. The inset plot displays the zoomed-in structure. The galaxy distribution in light-green color highlights one of the filaments (Filament ID 1029 in \citet{2014Tempel}) studied in this work. The magenta and green lines enclose the UVIT FOVs observed in the present study. The red circle represents the boundary of the Bootes void while the pair of blue lines encloses the FOV observed by us to examine the properties of void galaxies \citep{2021Pandey}.   }
    \label{fig:cone}
\end{figure*}

This work is based on a proposed deep FUV survey of filamentary structure conducted by {\em AstroSat}/ Ultraviolet Imaging Telescope \citep[UVIT, ][]{2017Tandon}. The rectangular box in Figure~\ref{fig:cone} encloses the targeted filamentary structure ($z \sim$ 0.07). The UVIT provides high spatial resolution ($\approx$ three times better than {\em GALEX}) and enhanced sensitivity that would be utilized to identify recently star-forming galaxies inside filaments, which would otherwise remain undetected in previous UV imaging surveys. We have conducted a similar deep UVIT survey of the Bootes void \citep[red circle in Figure~\ref{fig:cone};][]{1981Kirshner}. The observations led to the detection of a few void galaxy candidates, which were not identified in {\em GALEX} survey. We reported a distinct blueward shift in color of void galaxies compared to a sample of local galaxies \citep{2021Pandey}. Unlike voids and galaxy clusters, representing the rarest and densest environment in the cosmic web, the filaments host intermediate environment density. A detailed understanding of star formation activity across different LSSs will be possible when enough studies explore these environments in FUV.

{Substantial evidence has been reported in support of {amplification of} the star formation activity in galaxies residing in filaments, e.g., \citet{Darvish_2014} uncovers the possibility of mild galaxy-galaxy interaction or efficient gas accretion through inter-filamentary channels in filament galaxies. Filaments provide an active environment to trap cold gas, assist its cooling, and enhance star formation activity \citep{2019Liao}. Utilizing the deep UVIT observations of filaments, we compute SFRs and compare the FUV and optical morphology of filament galaxies to uncover signs of interactions or gas accretion signatures. We also probe the stellar mass-assembly history of filament galaxies using FUV$-$optical colors in this work. The results are compared with galaxies residing in different LSSs.

}
 





This paper is organized as follows: Section~\ref{sec:data} describes the observational data used in the work and UVIT data reduction techniques followed, and Section~\ref{sec:sample} explains our methodology applied in extracting filament galaxies from UVIT observations. We derive the photometric and chemical properties of filament galaxies in Section~\ref{sec:properties}. Various aspects of FUV emission of the sample, such as morphology and FUV SFRs are discussed in Section~\ref{sec:fuvem}. In Section~\ref{sec:massa}, we examine the mass assembly of filament galaxies. Our findings are discussed in Section~\ref{sec:DC}. A standard $\Lambda$CDM cosmology with ${\Omega}_M$ = 0.3, ${\Omega}_{\Lambda}$ = 0.7, and H$_{0}$ = 70 km s$^{-1}$ Mpc$^{-1}$ is assumed in this paper.

\section{Data}
\label{sec:data}
\subsection{Observations}
We primarily use proposed {\em AstroSat}/UVIT data observed in BaF2 filter centered at ${\rm \lambda_e = 1541}$ \AA\ (Observation ID: A09\_127T01\_9000003816, A09\_127T03\_9000003820; PI: Divya Pandey). The total field-of-view (FOV) of the UVIT channel is 28$^{\prime}$ in diameter. The pixel scale of the reduced image {is} 0.417$^{\prime\prime}$, whereas the zero-point of the filter is $\sim$ 17.78 mag \citep{Tandonetal2020}.
Under the proposal, we observe a filamentary structure at $z$ $\sim$ 0.073. The observed structure is located close to one of the largest voids in the northern hemisphere, the Bootes void (See Figure~\ref{fig:cone}). We proposed to observe the structure in two FOVs with an exposure time of 7~kilo seconds each. The central coordinates of the two pointing are: $\alpha$ = 215.85$^\circ$, $\delta$ = 46.1$^\circ$ (FOV1) and $\alpha$ = 217.15$^\circ$, $\delta$ = 46.12$^\circ$ (FOV2). Figure~\ref{fig:cone} shows the direction of UVIT pointings on the cosmic web.

Apart from the UV survey, we use archival imaging and spectroscopic data from Sloan Digital Sky Survey Data Release 16 \citep[$SDSS$ DR 16,][]{2020Ahumada} and Galaxy Evolution Explorer \citep[{\em GALEX}, ][]{2007Morrisey} in this work.


\subsection{UVIT data reduction and analysis}
We obtain the raw Level 1 version of the observations from {\em AstroSat} data archive\footnote{\url{https://webapps.issdc.gov.in/astro_archive/archive/Home.jsp}}. The level 1 data is fed into CCDLAB \citep{2021Postama}. CCDLAB is an automated pipeline used for reducing the L1 version of UVIT data into science-ready images. The data reduction includes drift correction, flat-field correction and is further corrected for additional distortion and fixed noise pattern. The corrected images from different orbits are stacked together through the centroid alignment process to obtain a deep, science-ready image. CCDLAB removes frames containing large numbers of photon counts due to cosmic ray shower, leading to a reduction in the total exposure time of the image. The effective exposure time (t$_{\rm exp}$) of the final images was 5.7 kilo seconds and 6.3 kilo seconds. The world coordinate system (WCS) was assigned to the final images using {\tt Astroquery} \citep{2019Ginsburg} appended with CCDLAB. {\tt Astroquery} uses the GAIA \citep{2016Gaia} catalog to assign coordinates to each pixel in the final image. The mean error in the astrometry was $\approx$ 0.$^{\prime\prime}$28. Along with the science-ready images, we get corresponding exposure maps as an output from the CCDLAB.

We run source extractor \citep[][]{1996Bertin} for detecting sources observed in the two sets of FUV images. The background is detected in automated mode. The detection and analysis threshold is fixed at 1.5$\sigma$ for source extraction and photometry. We consider the minimum number of pixels required to classify an intensity distribution as a source (DETECT\_MINAREA) to be 9. The 3$\sigma$ limiting magnitudes for our observation in FOV 1 and FOV 2 are 24.79 mag and 25.07 mag, respectively. We detected a total of 3574 sources combining both the FOVs. We perform Petrosian photometry to measure the integrated magnitude of the detected sources. We expect to detect irregular and disky galaxies in our UVIT FOVs where the Petrosian apertures may recover nearly all flux from an exponential light distribution \citep{2001Blanton}.  

\section{Identifying filament galaxies}
\label{sec:sample}
Only a fraction of the detected sources in our FUV observation belong to the filaments. Hence, identifying filament galaxies is an imperative part of our study. Multiple catalogs employ various algorithms to find filaments and associated galaxies in the LSS \citep[][]{2014Tempel, 2022Carron, 2014Alpaslan}. We use a catalog given by \citet[][hereafter Tempel catalog]{2014Tempel} in our study. \citet{2014Tempel} employs the Bisous model to detect filaments in an extensive database of galaxies observed with {\it SDSS}.

{Bisous algorithm randomly distributes fixed-sized cylinders on a large-scale galaxy distribution. The probability of a cylinder corresponding to a filamentary structure is based on the galaxy density contrast inside and outside the cylinder. Multiple Markov Chain Monte Carlo (MCMC) simulations are run to establish a network of well-aligned cylinders of fixed radius, varying length, and orientation. Using MCMC simulation, a visit map is created, which gives the probability of a region belonging to the filamentary structure. The ridges of this visit map where the galaxy density is highest are considered as the filament spine. Tempel Catalog} provides a detailed list of filaments and constituent filament galaxies. The information regarding the distance from the nearest filament spine (D$_{\rm fil}$), comoving distance, integrated magnitudes, and optical luminosities for each filament galaxy are included in the catalog. {\citet{2014Tempel} apply density-dependent peculiar velocity sampling scheme \citep{fog2} to mitigate {\it finger-of-god} effect \citep{1978Tully} in the galaxy distances used for filament extraction. The projection effect may also impact the distances measured in this study. As a result, the D$_{\rm fil}$ parameter used in this study may represent the lower limit of the actual separation between galaxies and filament spine. 
}

\begin{figure*}
    \centering
    \includegraphics[width=0.59\linewidth]{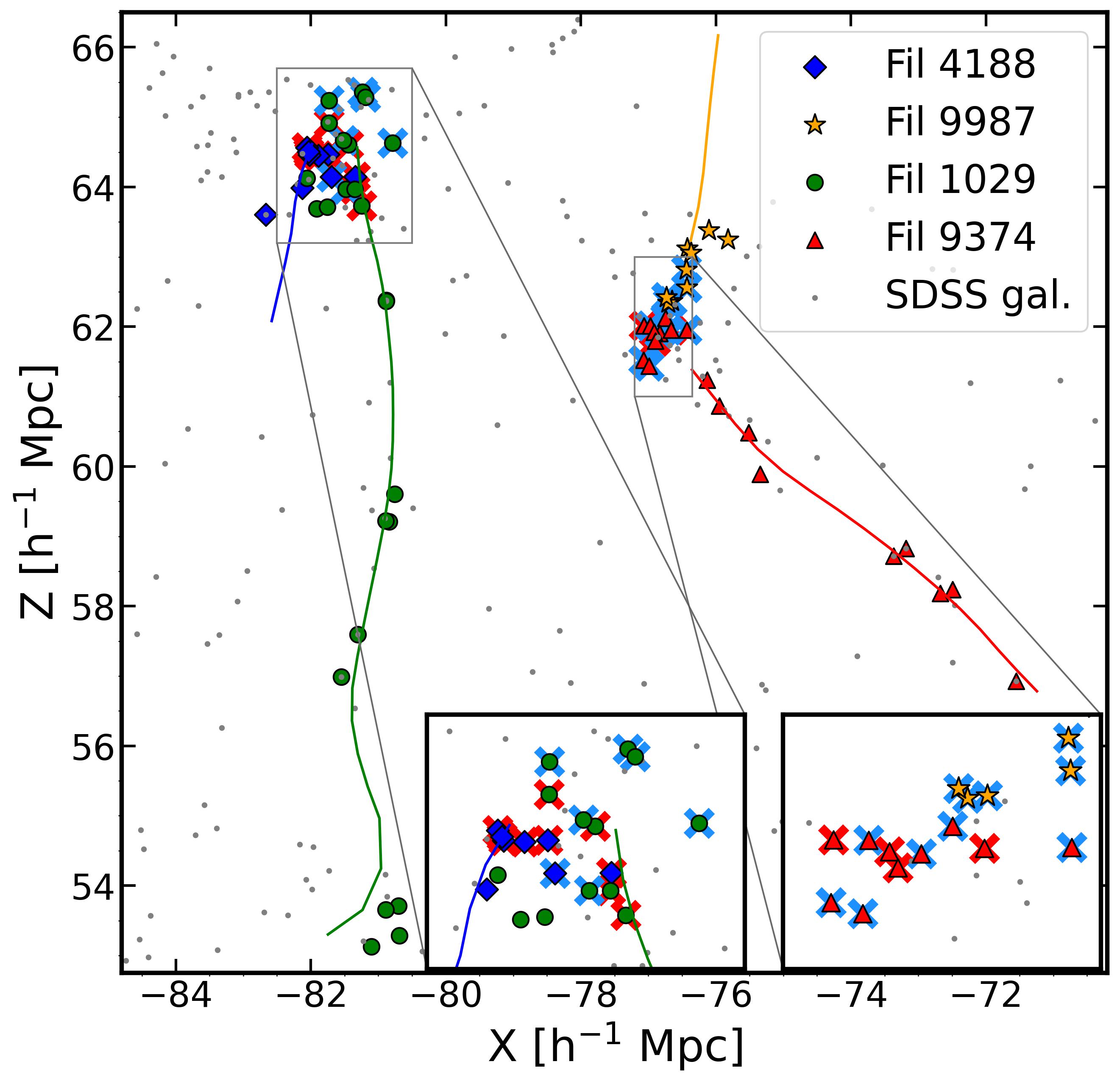}
    \caption{Sky map in a cartesian grid defined by \citet{2014Tempel} showing the distribution of galaxies in our FOV. Our UVIT pointings are enclosed in rectangular boxes. The galaxies within 1 Mpc $h^{-1}$ from the filament spine and the corresponding filament spine for each filament are shown with different symbols and colors. The detected galaxies in the FUV survey are outlined with blue crosses, whereas the red-colored crosses within FOVs represent filament galaxies that were FUV undetected. Grey markers show the background galaxies in the FOV from the {\em SDSS} survey.}
    \label{fig:skymap}
\end{figure*}

We matched our UVIT source catalog with the Tempel catalog within a radius of $\sim$2.$^{\prime\prime}$5 to identify filament galaxies in our FUV observations. The detected galaxies belong to four different filaments (Filament ID: 1029, 4188, 9987, and 9374 in the Tempel catalog) of different lengths. The four filaments together contain 58 galaxies, of which 18 were detected in the UVIT sample. {We restrict our filament sample up to a distance of 1 Mpc $h^{-1}$ from the filament spine (D$_{\rm fil}$), as previous report points that filaments considerably impact the properties of galaxies situated within 1 Mpc $h^{-1}$ from their axis \citep{2017Kuutma}}. A sky map depicting the distribution of galaxies in each filament and corresponding filament spine is shown in Figure~\ref{fig:skymap}. {The inset panels mark the position of UVIT pointings. Filament galaxies detected above the defined threshold are outlined with blue crosses, whereas those undetected are highlighted with red crosses.} Prior to UVIT, a shallow FUV survey (t$_{\rm exp}$ $\sim$ 100 s) of the observed FOVs was conducted by {\em GALEX}. We find that six of the 18 filament galaxies detected by UVIT remained missing in the GALEX AIS catalog. \textit{In other words, these filament galaxies are newly detected in our UVIT survey}. Table~\ref{tab:filament} consists of information on each filament, their constituent galaxies, and a comparison between filament galaxies detected by UVIT and GALEX AIS observation.


\begin{table}
 \caption{Details of the filament observed in our UVIT observation. }
    \centering
    \begin{tabular}{ccccc}
    \hline
    \hline
    {Fil. ID} & Length & Total Galaxies & FUV$_{\rm UVIT}$ & FUV$_{\rm GALEX}$\\
    (1) & (2) & (3) & (4) & (5) \\
    \hline
    9987 & 3 Mpc $h^{-1}$& 7 & 5 & 5\\
    9374 & 7.55 Mpc $h^{-1}$ & 19 & 6 & 5\\
    1029 & 13.69 Mpc $h^{-1}$ & 24 & 6 & 2 \\
    4188 & 3 Mpc $h^{-1}$ & 9 & 1 & 0\\
    \hline
    \end{tabular}
   \label{tab:filament}
 
    Col (1): Filament identity from \citet{2014Tempel}, Col (2): Length of each filament, Col (3): Total number of galaxies present in each filament within 1 Mpc $h^{-1}$ from the filament spine, Col (4): Number of galaxies detected in UVIT FUV and Col (5): Number of galaxies present in GALEX FUV catalog.
    
\end{table}

We extract Petrosian magnitudes corresponding to all {\em SDSS} broadband filters and spectroscopic information for all 58 galaxies from {\em SDSS} DR 16. The magnitudes used in this work are corrected for Galactic extinction using Galactic dust extinction maps provided by \citet{1998Schlegel}.

\section{Properties of filament galaxies}
\label{sec:properties}
\begin{figure*}
    \centering
    \includegraphics[width=0.99\linewidth]{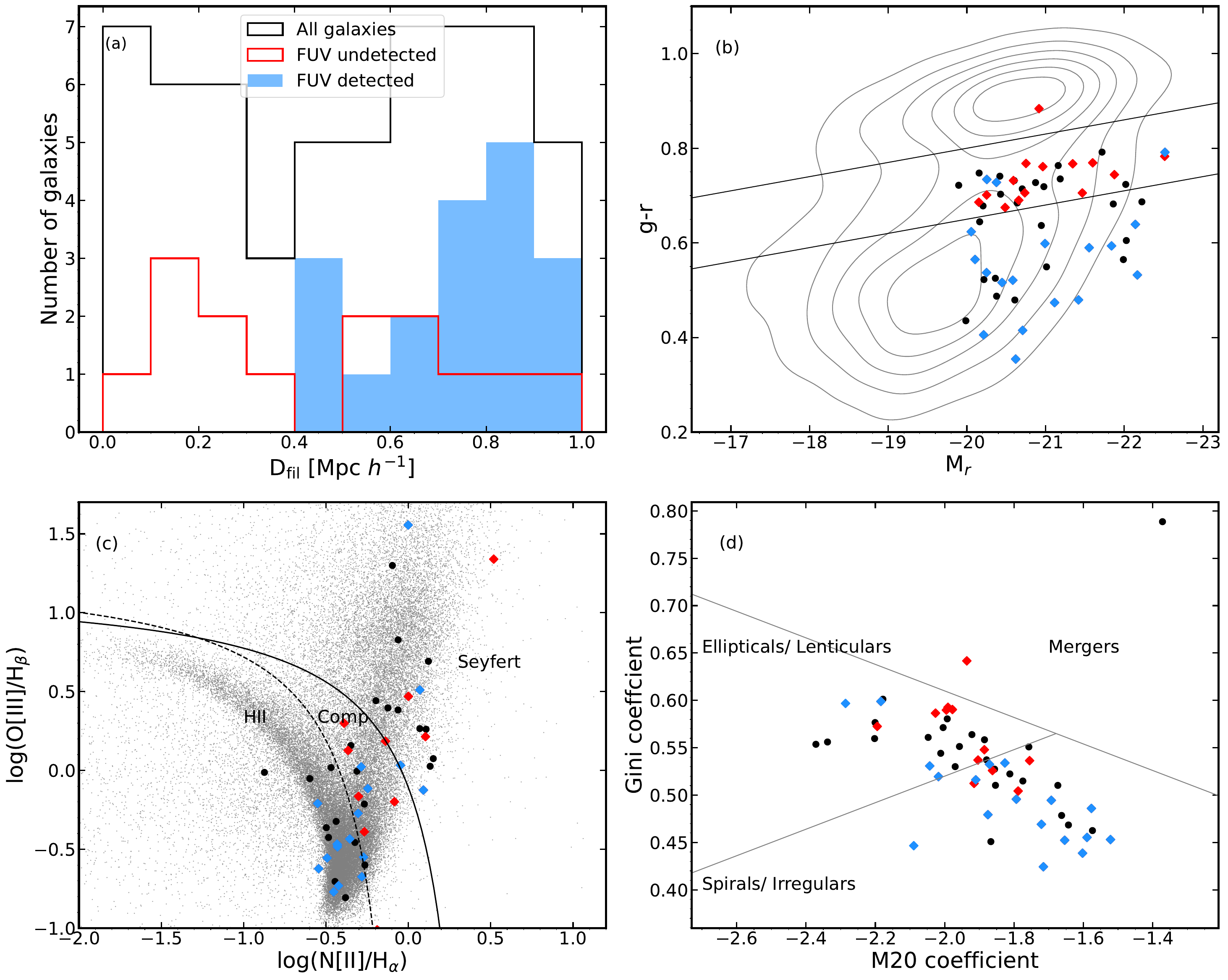}
    \caption{(a) Distribution of distance from the filament spine (D$_{\rm fil}$) for all filament galaxies (black) with D$_{\rm fil}$ $\leq$ 1 Mpc $h^{-1}$, galaxies detected (blue) and undetected (red) in our UVIT FUV survey. (b) g$-$r vs. M$_r$ color-magnitude diagram for filament galaxies. The contour is composed of a sample of local galaxies observed in {\em SDSS} DR 16. (c) BPT diagram used to separate HII region galaxies from composite and Seyfert galaxies. The solid and dashed polynomials are described in \citet{2001Kewley} and \citet{2003Kauffmann}, respectively. Each grey dot represents a galaxy present in \citet{2003bKauffmann} spectroscopic catalog. (d) Gini coefficient vs. M20 coefficient for all filament galaxies. The markers have the same meaning as (a).}
    \label{fig:combine_sec4}
\end{figure*}

We study the star-forming, nuclear, and structural properties of all 58 galaxies in the four filaments. As shown in Figure~\ref{fig:skymap}, only a portion of these filaments were observed with UVIT. Therefore, we create two subsets among the total galaxies observed with UVIT (within 1 Mpc $h^{-1}$ from the nearest filament axis): the FUV detected and FUV undetected galaxies. The histogram shown in the upper left panel of Figure~\ref{fig:combine_sec4} highlights the cross-sectional distribution of galaxies inside filaments from the filament axis for the entire sample and both subsets. The filaments seem to be more dense in the inner region. We emphasize that we detect no galaxy in our FUV observation up to D$_{\rm fil}$ $\approx$ 0.4 Mpc $h^{-1}$.  

\subsection{Color-magnitude diagram}
We use optical g$-$r vs. M$_r$ color-magnitude diagram (CMD) to classify all 58 filament galaxies into the blue cloud, red sequence, and green valley galaxies \citep{Blanton_2005, 2014Salim}. The blue cloud contains star-forming galaxies, mostly disk-dominated spirals, and irregulars. The red sequence comprises star formation quenched galaxies. Lenticular and elliptical galaxies mostly dominate this region. The intermediate region between the two modes of distribution is known as the green valley, where galaxies with intermediate star formation histories lie. The parallel lines dividing the CMD are given by \citet{Blanton_2005}. The background galaxies in the sample are taken from {\em SDSS} DR 16 within $z$ $<$ 0.1. 

As seen in the upper right panel of Figure~\ref{fig:combine_sec4}, our filament sample is spread across the blue cloud and green valley region. We detect a comparable population of star-forming galaxies and green valley galaxies in our FUV survey. On the contrary, we only detected star-forming and blue galaxies in our previous study of the Bootes void \citep{2021Pandey, 2022Pandey}. We highlight the dominance of green valley galaxies in filaments that contain a mix of stellar populations. Such galaxies are in a transition phase where the star formation is declining but not entirely ceased. Galaxy interactions and mergers, feedback from active galactic nuclei (AGN) are a few physical mechanisms that can drive a galaxy to this state \citep{2014Salim}. The galaxies present in filaments studied by us are brighter than M$_r$ $\sim$ $-$20 mag. However, previous reports have confirmed the presence of faint dwarf galaxies in filament environments as well \citep{2021Lee}.

\subsection{BPT diagram}
\label{sec:bpt}
The BPT \citep[Baldwin, Phillips, and Terlevich, ][]{1981Bpt} diagram is ubiquitously used to classify and understand the ionization sources of galaxies based on their emission line ratios. It provides insights into the dominant mechanisms that power the emission lines observed in the spectra of galaxies. We use the line flux ratio of [O III]/H$_\beta$ vs. [N II]/ H$_\alpha$. The diagram is divided into the three regions containing purely star-forming or `HII region' galaxies, composite galaxies, and galaxies with AGN based on prescription given by \citet{2001Kewley} and \citet{2003Kauffmann}.  

The emission line fluxes used in the diagram were corrected for internal extinction before use. The reddening $E(B-V)$ value was determined using H$_\alpha$ and H$_\beta$ emission lines following the Balmar decrement method given by \citet{1989Osterbrock}. It can be seen in the bottom left panel of Figure~\ref{fig:combine_sec4} that filaments contain a mix of populations in terms of nuclear activity. All purely star-forming galaxies in the FOVs were detected in FUV observation. We also detect a few galaxies with AGN in our UV survey. The AGN-host filament galaxies are star-marked in Table~\ref{tab:all18}.


\subsection{Morphological classification using non-parametric methods}
We employ a non-parametric approach to determine the morphological classification of filament galaxies studied in this work. This method requires us to compute the Gini and M20 coefficients. The Gini coefficient ($G$) is a statistical parameter based on the Lorentz curve \citep{Lotz_2004}. The parameter is computed with the following equation:
\begin{equation}
    G = \frac{1}{\Bar{X} n(n-1)}\sum^n_i(2i-n-1)X_i
\end{equation}
\noindent here, X$_i$ represents all pixel value , $\Bar{X}$ is the average of all X$_i$, $n$ is total number of pixels. All pixel values are arranged in an increasing order before calculating the $G$. For a homogeneous system, $G$ converges to 0, whereas $G$ = 1 when the entire light of a galaxy is concentrated in a pixel \citep{Gerald1962}. 

The parameter $M20$ corresponds to the second-order moment of light for the brightest region in the galaxy \citep{Lotz_2004}. The parameter is calculated by summing over the brightest pixel values ($f_i$) until the sum corresponds to 20\% of the total galaxy flux. This value is normalized by M$_{\rm tot}$ corresponding to the second-order moment of the total light in a galaxy. These quantities are defined as:

\begin{equation}
    M_{tot} = \sum_i^ n M_i = \sum f_i \big[(x_i - x_c)^2 + ( y_i - y_c)^2 \big]
    \label{mtot}
\end{equation}

\begin{equation}
    M_{20} = \log \Big ( \frac{\sum_i^ n M_i} {M_{tot}} \Big), {\rm where}\ \sum_i f_i < 0.2 f_{tot}  
    \label{m20}
\end{equation}

\noindent here, x$_c$, y$_c$ corresponds to galaxy central coordinates.


We use {\tt statmorph}, a python-based code provided by \citet{2019Rodriguez-Gomez} to calculate the above-defined statistical parameters for $SDSS$ $r$-band images of all filament galaxies. The code creates internal segmentation maps for the galaxies by using a threshold value for source detection and convolution filter size, both given by the user. The maps are used for calculating $G$ and $M20$. The galaxies whose fitting was suspected were not included in the analysis. 

The bottom right panel of Figure~\ref{fig:combine_sec4} shows the distribution for all filament galaxies. We refer to \citet{2008Lotz} for demarcating the galaxy population into spirals/ irregular, elliptical/ lenticular, and merger candidates on $G$ vs. $M20$ plane. The filament galaxies from our study are equally distributed in the late- and early-type galaxy regions. Only two clear cases of galaxy merger have been identified in the sample. These galaxies appear to have elliptical morphology. We primarily detect disky-spiral galaxies in our UVIT FUV survey. Filaments contain mixed galaxy populations compared to void galaxies, which are primarily blue and disky in morphology \citep{2012Kreckel, 2021Pandey}.

\begin{table*}
\caption{Detailed information on all 18 galaxies detected in UVIT FUV survey.}
\centering
\begin{tabular}{lcccccccr}
\hline
\hline
  \multicolumn{1}{c}{ID} &
  \multicolumn{1}{c}{RA} &
  \multicolumn{1}{c}{Dec} &
  \multicolumn{1}{c}{z} &
  \multicolumn{1}{c}{D$_{\rm fil}$} &
  \multicolumn{1}{c}{FUV}&
  \multicolumn{1}{c}{SFR$_{\rm FUV}$} &
  \multicolumn{1}{c}{log(M$_\star$)} &
  \multicolumn{1}{c}{Fil ID} \\
  & [deg] & [deg] & & [Mpc $h^{-1}$] & [mag] & [M$_\odot$ yr$^{-1}$] & [M$_\odot$] &  \\
 (1) & (2) & (3) & (4) & (5) & (6) & (7) & (8) & (9) \\
\hline 
  1 & 215.894 & 46.177 & 0.073 & 0.90 & 19.784 $\pm$ 0.042 & {1.502 $\pm$ 0.064} & 10.142 & 9987\\
  2 & 217.090 & 45.966 & 0.075 & 0.61 & 21.247 $\pm$ 0.089 & {0.531 $\pm$ 0.047} & 10.654 & 4188\\
  3 & 216.988 & 46.202 & 0.073 & 0.53 & 19.660 $\pm$ 0.043 & {1.239 $\pm$ 0.053} & 10.11 & 1029\\
  4 & 215.788 & 46.289 & 0.073 & 0.45 & 20.347 $\pm$ 0.056 & {1.238 $\pm$ 0.070} & 10.225 & 9987\\
  5* & 215.819 & 46.146 & 0.073 & 0.98 & 20.594 $\pm$ 0.074 & {4.841 $\pm$ 0.357} & 11.201 & 9987\\
  6 & 217.159 & 46.123 & 0.074 & 0.46 & 21.710 $\pm$ 0.102 & {0.547 $\pm$ 0.056} & 10.302 & 1029\\
  7 & 215.817 & 46.028 & 0.073 & 0.81 & 21.440 $\pm$ 0.091 & {0.405 $\pm$ 0.037} & 10.022 & 9374\\
  8 & 215.670 & 46.075 & 0.073 & 0.78 & 20.911 $\pm$ 0.08 & {0.638 $\pm$ 0.051} & 10.257 & 9374\\
  9 & 215.772 & 46.149 & 0.074 & 1.00 & 19.649 $\pm$ 0.043 & {5.140 $\pm$ 0.222} & 10.776 & 9987\\
  10 & 215.903 & 46.111 & 0.072 & 0.85 & 19.580 $\pm$ 0.041 & {3.071 $\pm$ 0.126} & 11.004 & 9374\\
  11 & 217.302 & 46.240 & 0.075 & 0.85 & 20.410 $\pm$ 0.078 & {0.527 $\pm$ 0.041} & 9.911 & 1029\\
  12 & 215.947 & 46.058 & 0.072 & 0.88 & 21.175 $\pm$ 0.083 & {0.625 $\pm$ 0.052} & 10.109 & 9374\\
  13 & 217.152 & 46.317 & 0.075 & 0.88 & 19.206 $\pm$ 0.037 & {2.162 $\pm$ 0.080} & 10.585 & 1029\\
  14 & 217.145 & 46.307 & 0.074 & 0.78 & 23.200 $\pm$ 0.256 & {0.050 $\pm$ 0.013} & 10.284 & 1029\\
  15 & 215.838 & 46.247 & 0.072 & 0.74 & 19.629 $\pm$ 0.046 & {1.883 $\pm$ 0.087} & 10.496 & 9987\\
  16* & 215.941 & 45.931 & 0.072 & 0.74 & 18.890 $\pm$ 0.034 & {1.867 $\pm$ 0.063} & 11.368 & 9374\\
  17* & 217.108 & 45.969 & 0.074 & 0.43 & 22.755 $\pm$ 0.175 & {0.036 $\pm$ 0.006} & 11.406 & 1029\\
18* & 215.848 & 45.905 & 0.072 & 0.67 & 24.893 $\pm$ 0.466 & {0.005 $\pm$ 0.002} & 10.301 & 9374\\
\hline\end{tabular}\\
{Col. (1) Identity number of each galaxy; (2), (3), and (4) represent the coordinates and redshift; (5) distance of each galaxy from the nearest filament spine; (6) FUV magnitudes (uncorrected for internal and galactic extinction); (7) intrinsic FUV SFRs; (8) stellar mass; (9) Serial number of the parent filament taken from \citet{2014Tempel}. The star-marked galaxies in the table host an AGN.}
\label{tab:all18}
\end{table*}

\section{FUV emission in filament galaxies}
\label{sec:fuvem}

\begin{figure*}
    \centering
    \includegraphics[width=0.49\linewidth]{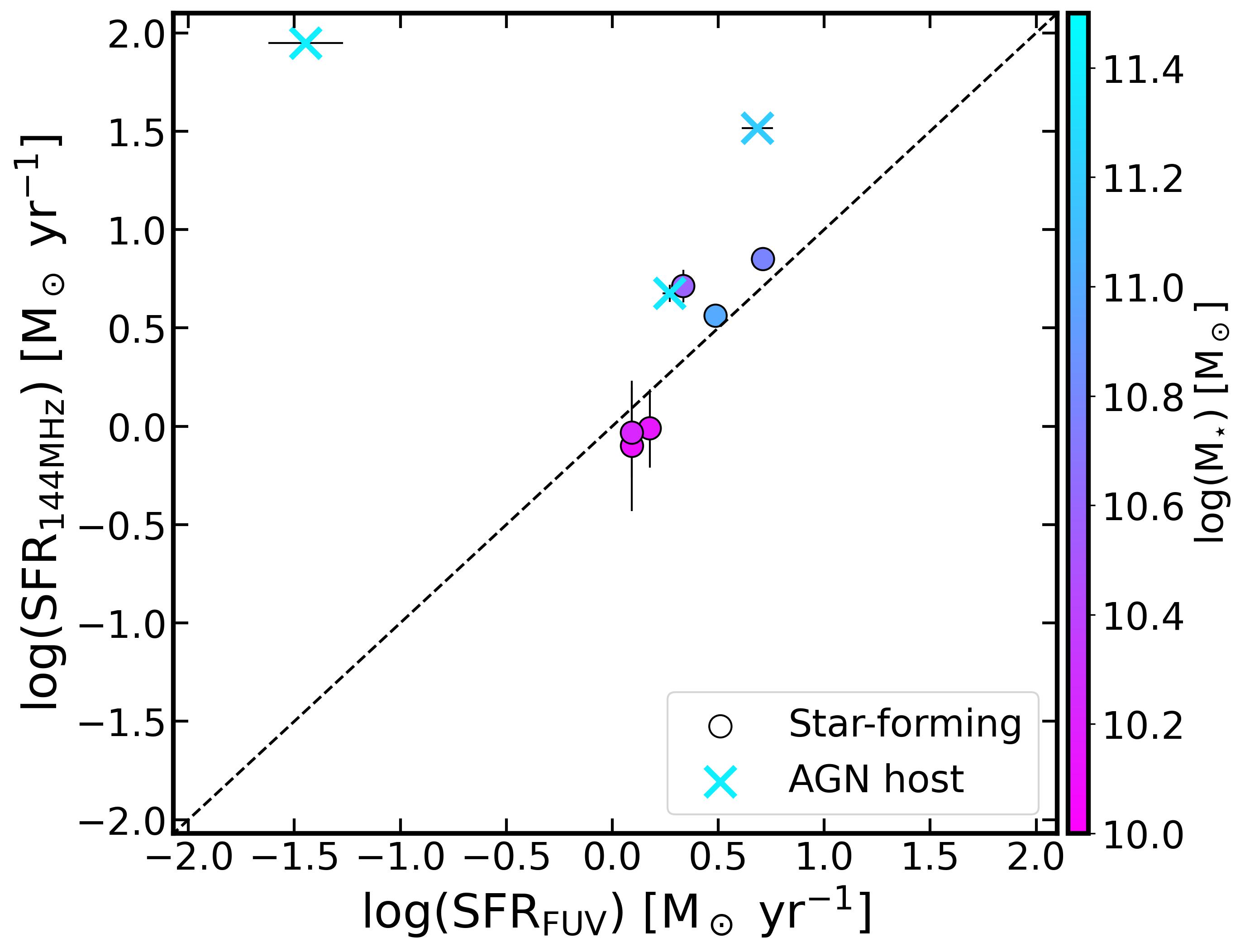}
     \includegraphics[width=0.49\linewidth]{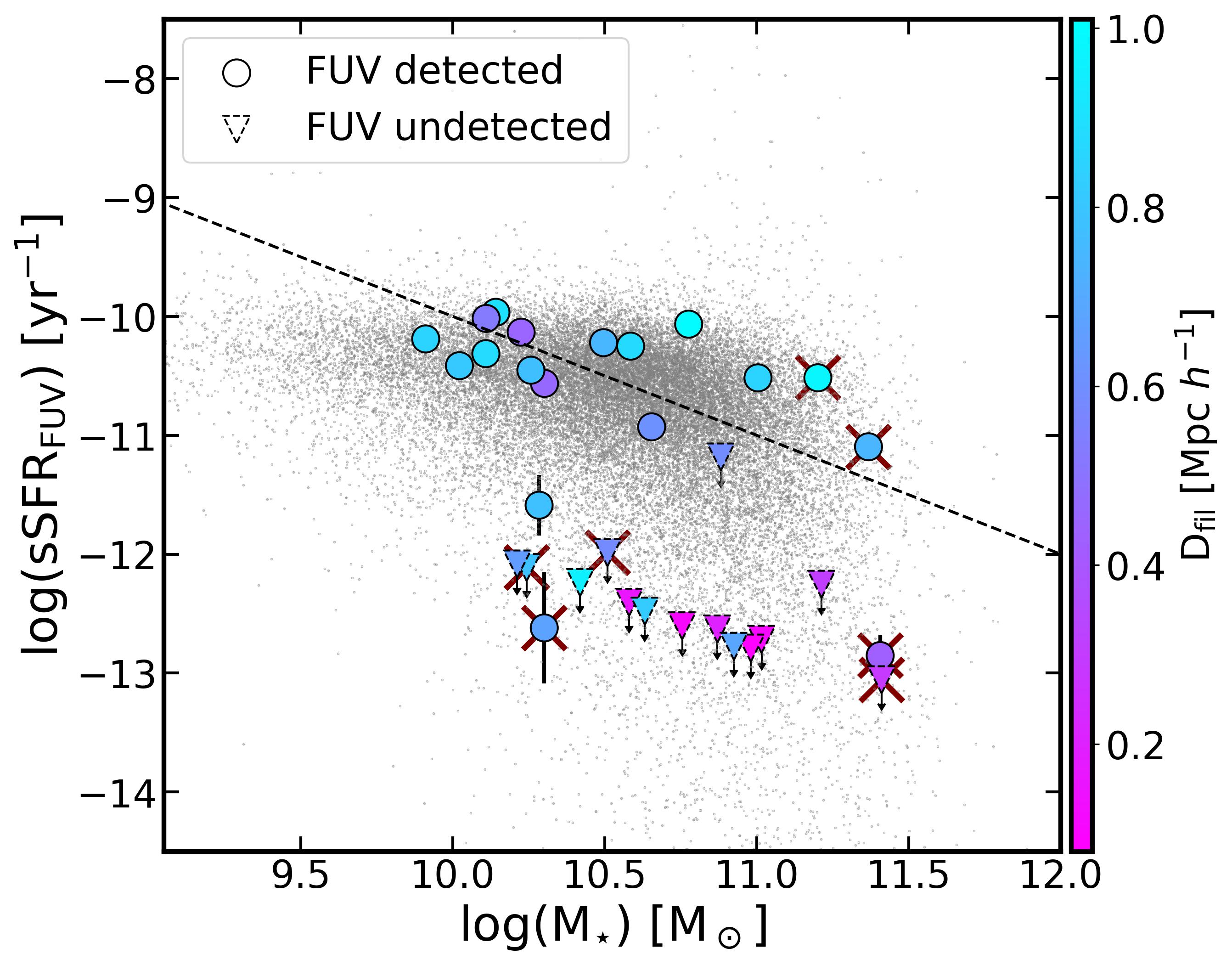}
    
    \caption{{Left Panel: Comparison between SFR$_{\rm FUV}$ and SFR$_{\rm 144 MHz}$ for star-forming (circles) and AGN (crosses) filament galaxies. The color bar represents M$_\star$. The dashed line indicates one-to-one correspondence between the quantities. Right Panel: sSFR vs. M$_\star$ for UVIT detected (circles) and undetected (triangles) filament galaxies. The downward triangular markers indicate the upper limit of sSFR$_{\rm FUV}$. The AGN-hosts are outlined with maroon crosses. The dashed line represents galaxies with SFR $=$ 1 M$_\odot$ yr$^{-1}$. The color bar represents the D$_{\rm fil}$ for each galaxy. Background galaxies shown with grey markers are taken from \citet{2016Salim}.}}
    \label{fig:fuvssfr}
\end{figure*}

The FUV SFRs would serve as an excellent tool to identify the most recent star-forming galaxy in our sample. As highlighted in the previous section, apart from UVIT-detected galaxies, we find a set of filament galaxies that do not emit in FUV above a threshold of 1.5$\sigma$. These galaxies are termed as FUV undetected. We perform fixed aperture photometry with an aperture radius of $\approx$ 1.5$^{\prime\prime}$ on the FUV images at the centroid positions of the undetected galaxies. The magnitudes, thus obtained, were aperture corrected up to the saturation radius of a point source. The average aperture corrected magnitudes were approximately 1 - 1.5 mag higher than the FUV 3$\sigma$ limiting magnitude. These magnitudes are used in the subsequent analysis in this work.

We calculated internal extinction corrected FUV SFRs for all (UVIT detected and undetected) filament galaxies using the following relation taken from {\citet{2012Kennicutt}}:
\begin{equation}
{\bf \rm \log SFR (M_\odot yr^{-1}) = \log L_{FUV} - 43.35 }
      \label{eq:sfr}
\end{equation}

where L$_{\rm FUV}$is the intrinsic FUV luminosity. We determine the value of internal extinction by the method discussed earlier in section~\ref{sec:bpt}. The ratio between stellar-to-gas dust attenuation is assumed to be 0.44 {\citep{calzetti1997aip}. Note that the stellar dust attenuation is affected by large-scale dust geometry in the case of galaxies hosting massive star-forming clumps. For such galaxies, the stellar-to-gas dust attenuation ratio may rise. Hence, a deviation from the local dust attenuation relation is expected. In other words, using the aforementioned local dust calibration relation, we may underestimate SFR from galaxies with the most clumpy FUV emission.} The median FUV SFRs of UVIT detected (non-AGN) filament galaxies is $\approx$ {1.4} M$_\odot$\ yr$^{-1}$, similar to the Milky Way galaxy. The stellar mass of filament galaxies was procured from GALEX-{\em SDSS}-WISE Legacy Catalog \citep[GSWLC,][]{2016Salim}. Table~\ref{tab:all18} contains the FUV magnitudes, SFRs, stellar mass, and D$_{\rm fil}$ for all filament galaxies detected in UVIT.

\subsection{SFR$_{FUV}$ $-$ SFR$_{\rm 144 MHz}$ relation}
{As discussed above, the FUV emission is subjected to high dust attenuation due to the presence of dust clouds in the interstellar medium (ISM). The absorbed emission is re-emitted in Infrared (IR) wavebands ($\lambda$ between 8 $\mu$m and 1000 $\mu$m), hence, a combination of FUV and IR fluxes gives a precise estimate of the SFR in galaxies \citep{2012Kennicutt, 2019Buat}. However, deep and well-resolved IR data is unavailable for our UVIT observed fields. The radio emission from star-forming galaxies is a promising alternative to estimate recent SFR. The emission at this wavelength in the star-forming galaxies is attributed to synchrotron radiation from accelerated electrons by the supernovae remnants, which is the end stage of short-lived massive stars and free-free emission from ionized gas around massive stars \citep{1992Condon, 1998Cram}. Most importantly, this emission is not impacted by dust attenuation. We use low-frequency radio continuum observation (120 - 168 MHz) from LOw-Frequency ARray (LOFAR) Two-meter Sky Survey DR 2 \citep[LoTSS;][]{2022Shimwell} to estimate SFR$_{\rm 144 MHz}$ for the FUV-detected filament galaxies. The survey has achieved a sensitivity upto $\sim$100 $\mu$Jy and angular resolution of 6$^{\prime\prime}$. The relationship between radio luminosity (L$_{\rm 144 MHz}$) and SFR at a short timescales ($\lesssim$ 100 Myr) for a star-forming galaxy is well-established \citep{2018Gurkan, 2021Smith}. 

We detected nine filament galaxies in the LOFAR survey, of which three host AGN. The L$_{\rm 144 MHz}$ (in W Hz$^{-1}$) for our sample were calculated using total flux density (mJy) from LOFAR source catalog assuming a spectral slope $(\alpha)\, = 0.7$ \citep{2016Hardcastle}. The L$_{\rm 144\, MHz}$ is calibrated to SFR$_{\rm 144\, MHz}$ following \citet{2017Calistro}. The distribution of SFR$_{\rm FUV}$ and SFR$_{\rm 144\, MHz}$ for star-forming (circles) and AGN-host (crosses) filament galaxies is shown in the left panel of Figure~\ref{fig:fuvssfr}. The color bar in the diagram denotes stellar mass. The non-AGN host galaxies with SFR$_{\rm FUV}\, <\, 1$ M$_\odot$ yr$^{-1}$ from our sample were not identified in the LOFAR catalog. We highlight that both SFR indicators are in close agreement with each other for star-forming galaxies, while SFR$_{\rm 144 MHz}$ in active galaxies is exorbitantly high compared to its FUV counterpart. The radio emission in such galaxies is powered by synchrotron emission from accreting central black holes.}

\subsection{sSFR$_{\rm FUV}$ versus M$_\star$ as function of D$_{\rm fil}$}

{The star-forming properties of galaxies depend on multiple factors, such as the halo-spin alignment of galaxies and D$_{\rm fil}$ \citep{2017Kuutma,2018mahajan,2018GaneshaiahVeena,2021GaneshaiahVeena}. In this section, we discuss the variation in star-forming properties of galaxies with D$_{\rm fil}$. The cross-sectional region of a filament is composed of three zones. The innermost `stream zone' channelizes cold gas to massive galaxies, followed by the `vortex zone' where the velocity is dominated by quadrupolar vortex structure risen due to the inflow of gas through sheets towards the center of filament, and the outermost `thermal zone' where inflowing gas is decelerated due to thermal pressure \citep{2024Lumandelkar}. Studies investigating the connection between star formation quenching in the LSS have produced mixed results. While some reports suggest the cessation of star formation near the filamentary structures \citep{2017Kuutma,2018Kraljic, 2020Singh}, a few observations report enhancement in star formation near filaments \citep{2017Kleiner,2019Vulcani}. }

The right panel of Figure~\ref{fig:fuvssfr} shows FUV specific-SFR {(log(sSFR) $=$ log(SFR) $-$ log(M$_\star$))} vs. M$_\star$ distribution for all filament galaxies targeted in our UVIT survey. We use aperture-corrected FUV magnitudes to calculate FUV SFRs in the case of UVIT undetected galaxies. The color bar indicates the D$_{\rm fil}$ for each galaxy. {The AGN-hosts in our sample are marked with a maroon cross in the figure. These galaxies are removed from the analysis hereafter.}  

As seen in the figure, we find two distinct populations of galaxies in the filaments -- star-forming and quiescent. A majority of the UVIT undetected galaxies lie closer to the filament spine. The non-detection of galaxies near the spine could be associated with the extra-galactic gas supply cutoff deep inside filaments suggested by \citet{2017Kuutma}, which may lead to { decline in recent star formation. We find that sSFR$_{\rm FUV}$ and D$_{\rm fil}$ positively correlate for our combined sample. In other words, sSFR in galaxies increases with an increase in D$_{\rm fil}$.}

Notably, most intermediate and high stellar mass galaxies detected in our survey are star-forming. It is well-established that several stellar mass quenching processes, namely gravitation quenching \citep{Genzel_2014}, AGN feedback \citep{2006Croton}, active in massive galaxies lead to a decline in the ongoing star formation. A few studies on filaments establish that the inter-filamentary gas can replenish the H{\sc I} content of filament galaxies \citep{2005Keres}. \citet{2017Kleiner} reported the evidence of cold-mode gas accretion in several galaxies. This effect is more prominently observed in high-mass galaxies due to their deep gravitational potential well. The accumulated cold gas can further ignite the star formation activities in the massive filament galaxies. However, the likelihood of gas accretion in the sampled filament galaxies needs further investigation.





\subsection{Comparison with galaxies from different LSSs}

\begin{figure}
    \centering
    \includegraphics[width=1.0\linewidth]{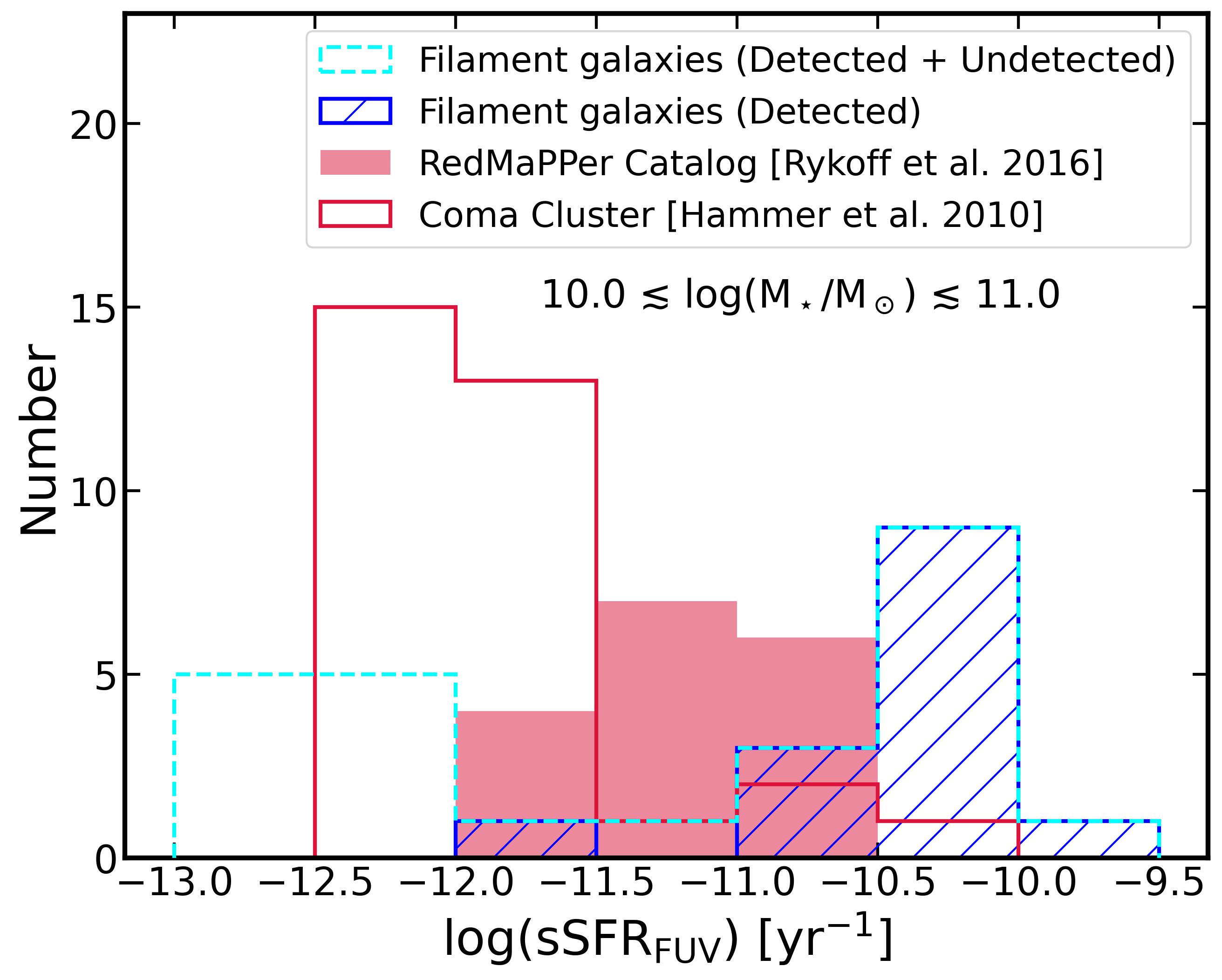}
    \caption{Histogram showing distribution of log(sSFR$_{\rm FUV}$) for detected/ detected$+$undetected filament galaxies (hatched blue/dashed cyan), a sample of cluster galaxies from RedMaPPer catalog \citep[filled red;][]{2016Rykoff} and Coma cluster \citep[unfilled red;][]{2010Hammer}. The galaxies fall in the same stellar mass range.}
    \label{fig:cluster}
\end{figure}

One of the key objectives of this work is to investigate whether the star-forming properties of filament galaxies are different from those existing in other LSSs, namely the voids and clusters. Most of the (non-AGN) galaxies detected in our FUV survey lie on the massive side of the stellar mass range (10$^{9.9}$ M$_\odot\, \leq$ M$_\star\, \leq$ 10$^{11.}$ M$_\odot$) while the median log(sSFR$_{\rm FUV}$) $\sim\, -10.4$ yr$^{-1}$  for filament galaxies. {For a small sample of void galaxies between stellar mass ($\sim$10$^{10}$  M$_\odot$ to 10$^{10.5}$  M$_\odot$) taken from Void Galaxy Survey \citep[VGS;][]{2012Kreckel}, the log(sSFR) $\approx$ $-9.82$ yr$^{-1}$, higher than our filament sample. Similarly, \citet{2024Conrado} report slightly elevated sSFRs in void galaxies compared to wall and filament galaxies. }


Galaxy clusters are another LSS known to host the most massive galaxies in the Universe. We compare sSFR for filament and cluster galaxies in a fixed stellar mass bin (10$^{10.0}$ M$_\odot\, \lesssim$ M$_\star\, \lesssim$ 10$^{11.0}$ M$_\odot$). We use the RedMaPPer catalog \citep{2016Rykoff} for identifying cluster galaxies in the nearby Universe (z $<$ 0.1). We select galaxies with cluster membership probability greater than 50\%. The FUV magnitudes of cluster members were obtained from the {\em GALEX} AIS catalog. The galaxies with FUV magnitude error less than 0.35 mag are included in the sample. After applying all the cuts, we obtain a small sample of 18 (non-AGN) cluster galaxies. Using the FUV magnitudes, we calculate internal dust extinction FUV SFR using Equation~\ref{eq:sfr}. The value of internal reddening in the galaxies is deduced using the Balmar decrement ratio method (see Section~\ref{sec:bpt}). {We create another dataset of cluster galaxies using deep {\em GALEX} observation of Coma cluster \citep{2010Hammer}. The observations reach a detection limit of 25 mag in FUV, equivalent to our UVIT observation. We identify cluster galaxies from {\em GALEX} observation following the structure of the Coma cluster defined in \citet{1996Colless}. The FUV magnitudes were corrected for internal dust extinction using A$_{\rm FUV}$ = 1.25$\pm$0.25 mag \citep{2008cortese} before SFR estimation. The stellar mass corresponding to both set of cluster galaxies were taken from GSWLC. We separate purely star-forming galaxies and AGNs using the BPT diagram from both sets and apply the stellar mass limit.  


As shown in Figure~\ref{fig:cluster}, we find that filament galaxies have higher values of sSFR than both sets of cluster galaxies, implying that the filament galaxies are more star-forming than their counterparts in clusters.} The observed results could arise from a slower gas depletion time scale in the former environment than the latter one, as pointed out by \citet{2022Castignani}. Furthermore, the cold mode gas accretion from inter-filamentary channels or gas-rich mergers could rejuvenate the star formation in massive galaxies. {It is worth noting that \citet{2024Parente} presented a similar trend for galaxies with M$_\star$ $\leq$ 10$^{10.8}$ M$_\odot$, showing higher sSFR with decreasing environment density; the star formation remains unaffected by the environment for more massive systems.   }

\subsection{FUV morphology of filament galaxies}
\label{sec:fuvm}

\begin{figure*}
    \centering
    \includegraphics[width=1.05\linewidth]{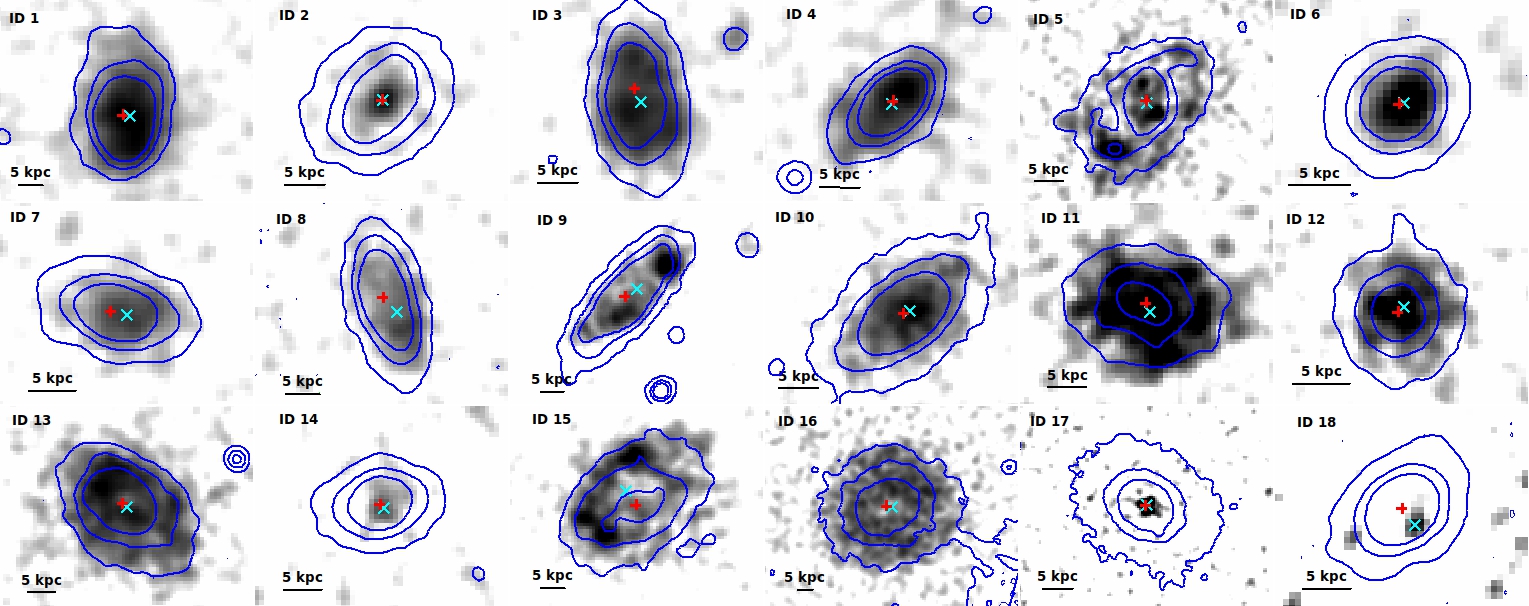}
    \caption{Grey-scale FUV image stamps of all 18 galaxies detected in the UVIT survey are shown. The optical morphology in {\em SDSS} r-band is superimposed on the image stamps. The three levels of the blue contours represent 10$\sigma$, 5$\sigma$, and 1.5$\sigma$ threshold level above the background noise. The surface brightness of the outermost contour is 24.14 mag arcsec$^{-2}$. The identity numbers mentioned on each image correspond to Table~\ref{tab:all18}. We mark the centroids of FUV and r-band observation with cyan and red crosses, respectively.}
    \label{fig:fuvr}
\end{figure*}

The FUV morphology of galaxies is an efficient probe to identify the clumpy star-forming regions inside them. We utilize our well-resolved UVIT observation to investigate the FUV morphology of filament galaxies and compare them with their optical counterparts. Figure~\ref{fig:fuvr} illustrates FUV images of all UVIT-detected filament galaxies. The blue contours on the figure show the {\em SDSS}-r band profile of galaxies. {The three contour levels in blue represent 10$\sigma$, 5$\sigma$, and 1.5$\sigma$ level, where $\sigma$ represents the mean standard deviation in the {\em SDSS} r-band observation. The surface brightness of the outermost contour is 24.14 mag arcsec$^{-2}$.} 
The galaxies are serially assigned a number ($ID$) according to Table~\ref{tab:all18}. 

{We show the FUV and optical centroids of the galaxies using cyan and red crosses, respectively.} The young stellar population inside galaxies represented by FUV emission mostly aligns with the main stellar population traced by optical emission. On the contrary, a similar analysis done for cluster galaxies by \citet{2022Mahajan} highlights that the FUV morphology for a sample of cluster galaxies was skewed relative to their optical emission primarily due to ram-pressure stripping \citep{1972Gunn} rampantly observed in the clusters. The result establishes that filament galaxies may not witness extreme environmental disturbances, which are prevalent in clusters. However, the FUV emission appears clumpy in the filament galaxies and shows considerable size variation. For a few galaxies, the emission is asymmetric within the galaxy, e.g., $ID\, 9, ID\, 15$. Moreover, we observe FUV emission in some galaxies beyond the outermost optical contour, e.g., $ID\, 1, ID\, 5, ID\, 11, ID\, 15$. These observations hint at the possibility of recent gas accretion episodes. {A thorough comparison between deep optical and UV observation of the filament galaxies is required for confirming the existence of extended-UV disks \citep{Thilker_2007,Borgohainetal2022}. }

As discussed in the previous section, massive galaxies in filament exhibit {slower star formation quenching} than those in the clusters. The FUV disk sizes of most of the filament galaxies in the sample are slightly extended or comparable to their optical emission. These galaxies are mostly detected in the outer edge of filaments (D$_{\rm fil}\, \gtrsim 0.7$ Mpc $h^{-1}$). Previous studies establish that massive and older galaxies reside closer to the filament center \citep{2017Chen, 2017Kuutma}. In our survey, massive galaxies near the filament axis were undetected or detected with weak UV emission (see $ID\, 17$ in Figure~\ref{fig:fuvr}; also see upper left panel of Figure~\ref{fig:combine_sec4}). An earlier study analyzing the HI gas content of galaxies in the filament environment reported that at fixed stellar mass and color, the galaxies closer to the filament spine are increasingly HI deficient \citep{Odekon_2018}. This finding corroborates our result that the galaxies closer to the spine do not show a signature of recent star formation.


\section{Impact on the mass assembly}
\label{sec:massa}

The results presented in the earlier sections establish that the filaments provide an in-homogeneous environment for galaxies to evolve. We detect negative gradients in FUV sSFRs on approaching the filament spine. Moreover, we find hints of gas accretion on the filament galaxies by comparing their UV and optical morphology. The observations suggest the existence of physical mechanisms, such as galaxy interaction and gas starvation deep inside filaments. The study of the stellar mass assembly of filament galaxies would highlight the mechanisms dominantly influencing their evolution. Multiple physical in-situ and externally driven processes may lead to stellar mass accumulation in galaxies over the cosmic time. The in-situ process includes star formation, gas recycling, and gas accretion from the inter-galactic medium \citep{1980Fall, carroll_ostlie_2017} whereas merger-interaction, accretion of satellite galaxies, gas stripping \citep{toomre1977evolution} are a few externally driven processes that steer stellar mass growth. Two prominent mass assembly mechanisms are considered in action in galaxies, i.e., `inside-out' and `outside-in' modes in the literature \citep{sanchez2007spatially, 2013Perez, Pan_2015}. In the case of inside-out mass assembly, the central regions of a galaxy form stars earlier and more rapidly than the outer regions. Outside-in growth, on the other hand, describes a pattern where star formation starts in the outer regions of a galaxy and then progresses inward toward the central region. The secularly evolving galaxies mainly grow via an inside-out mechanism, whereas galaxies influenced by the external environment show outside-in stellar mass build-up. 

{There are many excellent probes to investigate the stellar mass assembly of galaxies, e.g., a combination of well-resolved near-IR and H$\alpha$ observations which trace long-term and very recent ($\sim$ 10 Myr) star formation, respectively, would be effective in tracing the influence of gas accretion and longer-term inside-out star formation history. However, we lack such observations for our sample.} Broadband colors of galaxies are another alternative often used to determine the stellar population age of galaxies. This section uses the UV-optical ($FUV-r$) color index to indicate stellar population age. The ($FUV-r$) color is used actively to segregate star-forming and passive galaxies \citep{2012Hammer}. Both {\em SDSS} and UVIT have similar resolution and pixel scale ($\sim$ 0.${^{\prime\prime}}$4). We perform photometry on a central aperture of r = 1.5$^{\prime\prime}$ as it is roughly comparable to the point-source function full-width half maxima (PSF FWHM) of the two surveys. The radius is equivalent to $\approx$ 2.25 kpc at $z$ between 0.07 and 0.08. {Note that AGNs are excluded in this analysis.}

We consider the Petrosian magnitude of the sources as a measure of their total magnitude. The color gradient, $\Delta (FUV-r)$, is computed using the following equation:  
\begin{equation}
    \Delta(FUV-r)= (FUV-r)_{r=1.5^{\prime\prime}} - (FUV-r)_{\rm total}
    \label{eq:colorgrad}
\end{equation}

\noindent We divide our galaxy sample into three categories based on the value of $\Delta (FUV-r)$. The color gradient is positive ($> 2\sigma$) for red-cored galaxies whereas $\Delta (FUV-r)\, <\, 2\sigma$ for blue-cored galaxies. The galaxies with $|\Delta (FUV-r)| \, <\, 2\sigma$ are considered flat-cored. The average error ($\sigma$) associated with (FUV$-$r) color is 0.12 mag.


\begin{figure*}
    \centering
   
    \includegraphics[width=0.49\linewidth]{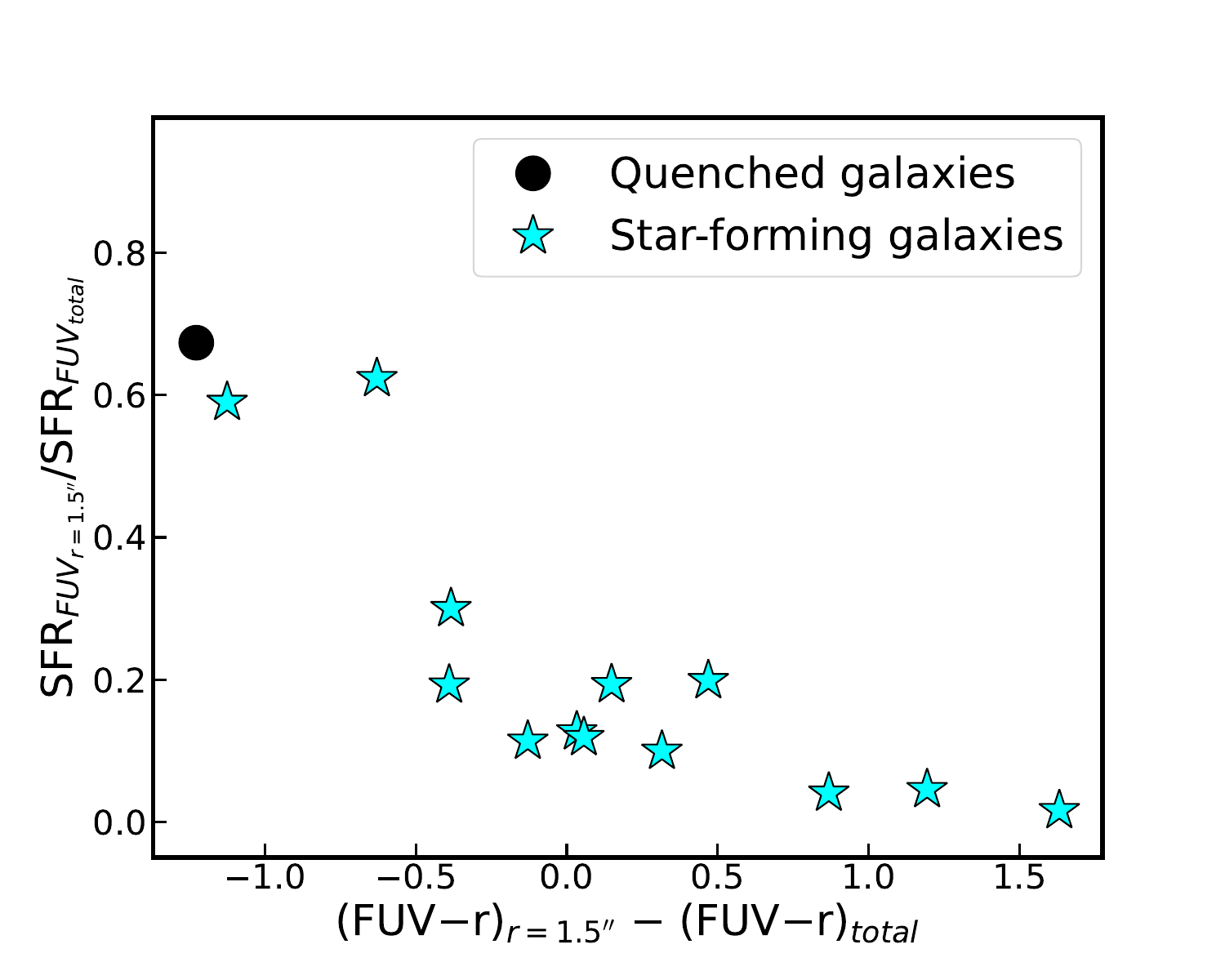}
    \includegraphics[width=0.49\linewidth]{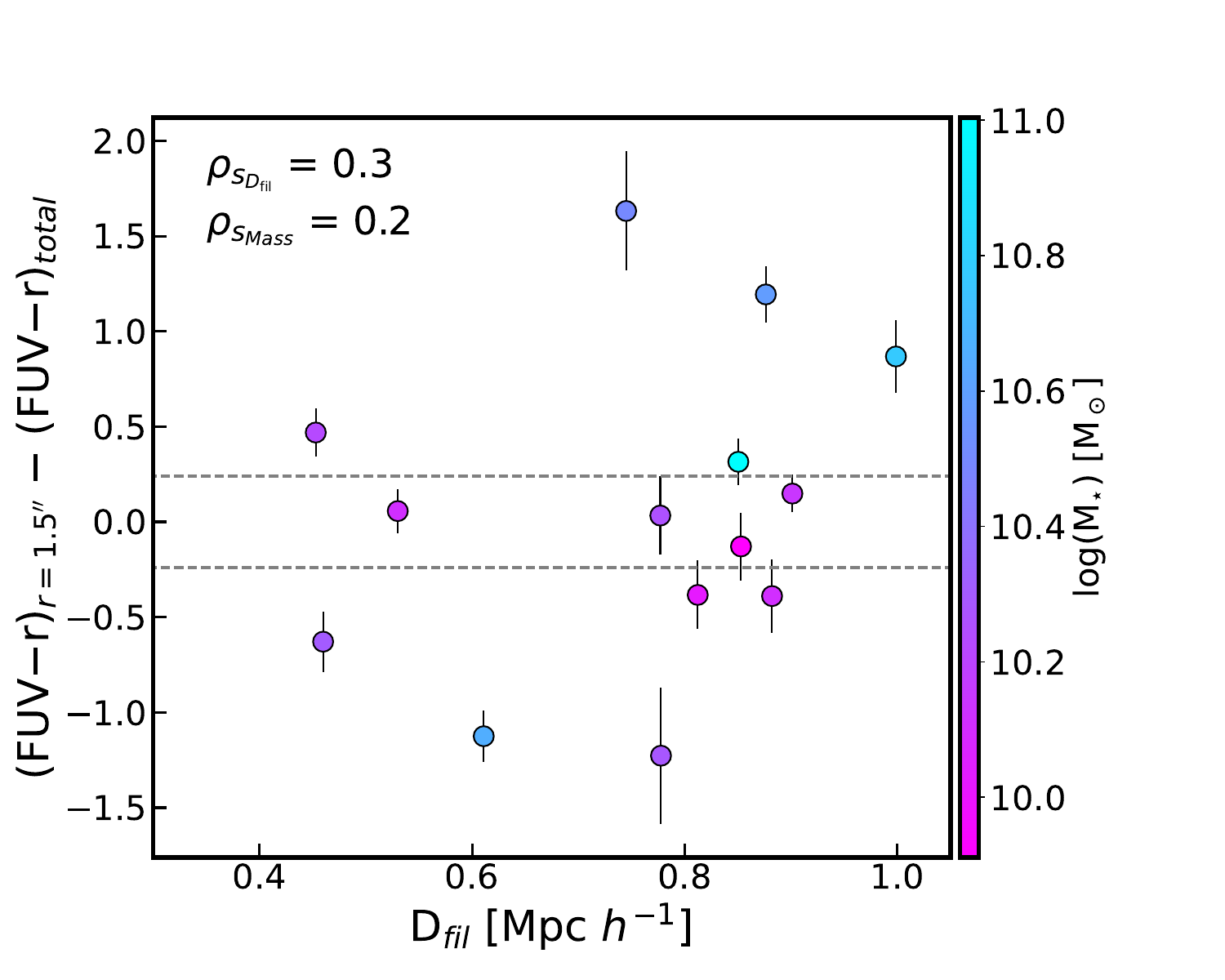}
    \caption{Left panel: Ratio of FUV SFRs (aperture radius = 1.5$^{"}$ / Petrosian aperture) versus $\Delta (FUV-r)$. Right panel: Distribution of $\Delta (FUV-r)$ and D$_{\rm fil}$. The color bar shown in the image represents the stellar mass of individual filament galaxies. The horizontal lines enclose the region representing flat-color gradient galaxies. }
    \label{fig:colorgrad}
\end{figure*}

We calculate the ratio of FUV SFR within a circular aperture of radius $\sim$1.$^{\prime\prime}$5 to total FUV SFR. The left panel of Figure~\ref{fig:colorgrad} shows the central-to-total SFR ratio vs. $\Delta (FUV-r)$ distribution for filament galaxies. Here, galaxies belonging to the blue cloud in the upper right panel of Figure~\ref{fig:combine_sec4} are termed star-forming, whereas the remaining sample detected in UVIT is grouped as quenched. The quenched galaxy in the sample has the most negative color gradient, similar to blue-cored galaxies. The star formation in such galaxies is restricted to the inner region. On the other hand, the red-cored filament galaxies have comparable FUV SFR throughout their extent.

The variation of $\Delta (FUV-r)$ and D$_{\rm fil}$ is shown in the right panel of Figure~\ref{fig:colorgrad}. Galaxies with M$_\star$ $<$ 10$^{10.5}$ M$_\odot$ mostly have flat color gradients with few exceptions showing a blue-core. A majority of blue-cored galaxies lie closer to the filament spine. The result implies that these galaxies are likelier to witness galaxy nurture activities. The galaxies situated radially outwards in the filament are red-cored or, in other words, follow `inside-out' mass assembly. We compare the strength of the monotonic relation between $\Delta (FUV-r)$ and D$_{\rm fil}$ against $\Delta (FUV-r)$ and M$_\star$ by computing their Spearman correlation coefficient ($\rho_s$) for both pairs. {We find that the quantities do not correlate significantly with $\Delta (FUV-r)$.}


\section{Discussion and Conclusions}
\label{sec:DC}
This study aims at understanding the star formation properties of galaxies inside filaments. We conduct a deep FUV survey to identify galaxies undergoing recent star formation and to subsequently investigate their FUV sSFRs, FUV morphology, and mass-assembly mechanism. The results are compared with cluster galaxies in a fixed stellar mass range. Although the filament comprises both early- and late-type galaxies, we mostly detect star-forming and disky galaxies in our FUV survey. Our UV survey does not detect galaxies closer than 0.4 Mpc $h^{-1}$ from the spine. Previous studies have pointed out that massive or red galaxies are more likely to exist near the spine of the filament \citep{2017Chen, 2017Kuutma}. The non-detection of such galaxies in our deep UV survey confirms the lack of star formation deep inside filaments. The mixed galaxy population type detected in filament could result from the pre-processing of galaxies inside filaments \citep{2022Castignani}.

Moreover, we find a positive trend between sSFR$_{\rm FUV}$ in galaxies with D$_{\rm fil}$. The galaxies at the filament outskirts are more star-forming than those in the inner region. The result corroborates the finding of \citet{2017Kuutma}, where the spiral galaxy fraction increases with increasing distance from the spine of the filament. The cluster galaxies show no variation SFR (or sSFR) across the cluster environment \citep{Lagana2017}. However, some studies report a drop in SFR with cluster-centric distances \citep{2004Balogh,2018mahajan}. In the case of void galaxies, \citet{ricciardelli} found a decrease in the SFR as the distance from the void center increases to a certain mean distance, while no effect could be established for the sSFRs of galaxies. The discussion above illustrates the difference in the evolution of galaxies in various LSSs. A comprehensive study combining more UV observations of LSSs would be aimed in the future.

In this study, we observe that a few intermediate-to-high stellar mass galaxies residing in the outer region of filaments have equivalent or slightly extended FUV disk sizes compared to their optical counterpart. The result may be an outcome of cold-mode gas accretion on the galaxies through filaments. However, some of these galaxies host AGN or are surrounded by close neighboring galaxies, which may have led to the observed result. Therefore, our study could neither confirm nor deny the cold mode accretion process reported by \citet{2017Kleiner}. Agreeably, we detected a limited number of filament galaxies in our UVIT survey. Hence, we aim to perform a similar study with an increased sample to enhance our understanding of filament galaxies. Deep multi-band observations could be utilized to uncover the prevalence of cold-mode gas accretion in filaments.

We study the mass assembly of filament galaxies based on $(FUV - r)$ color gradient. The galaxies closer to the filament spine mainly evolve via outside-in growth mass assembly. Similar results were reported by \citet{2021Lee}, where the high stellar mass galaxies closer to the filament spine show elevated H$_\alpha$ equivalent widths, indicating central star formation. A considerable cause of the central star formation could be the increased likelihood of merger interactions. {Our results largely depend on the estimated D$_{\rm fil}$, which may represent the lower limit of the actual values due to the projection effect.}

Our work establishes that filament could serve as a hotbed for several activities impacting the growth of galaxies residing inside filaments., e.g., galaxy interactions, gas starvation, and gas accretion. As a result, the filament galaxies display a wide range of galaxy properties in color, SFR, and morphology. The results from our study are summarized below:\\
i) We do not detect galaxies within D$_{\rm fil}$ $<$ 0.4 Mpc $h^{-1}$ in our deep UV survey.\\
ii) The study establishes that the filament environment substantially affects the evolution of galaxies inside them. We find early- and late-type galaxies residing inside the filament. Most of the galaxies detected in our FUV survey are star-forming and late-type. \\
iii) {The FUV SFR for star-forming filament galaxies is in agreement with SFR$_{\rm 144 MHz}$}.\\
iv) We witness an increase in FUV sSFR as the distance from the filament axis increases. \\
v) The filament galaxies show higher sSFR$_{\rm FUV}$ compared to cluster galaxies in a fixed same stellar mass range.\\
vi) The UV morphology for a few galaxies residing in the outer edge of filaments is extended more than their optical profile.\\
vii) `Inside-out' mass-assembly mode is more prevalent in the outer region of the filaments than in the inner region of the filament. The result could be due to an increased rate of galaxy merger interaction closer to the filament spine.

\begin{acknowledgments}
This publication uses the data from the AstroSat mission of the Indian Space Research Organisation (ISRO) archived at the Indian Space Science Data Centre (ISSDC). DP and ACP would like
to acknowledge the Inter University Centre for Astronomy and Astrophysics (IUCAA), Pune, India for providing facilities to carry out this work.
\end{acknowledgments}

%

\vspace{5mm}
\facilities{{\em AstroSat}/UVIT, {\em GALEX}, {\em SDSS}, LOFAR}


\software{CCDLAB \citep{2021Postama}, Source Extractor \citep{1996Bertin}, StatMorph \citep{2019Rodriguez-Gomez}
          }





\bibliography{sample631}{}
\bibliographystyle{aasjournal}



\end{document}